\documentclass[letter]{aa} 

\usepackage[utf8]{inputenc}
\usepackage[varg]{txfonts}
\usepackage[breaklinks, colorlinks, citecolor=blue, linkcolor=blue]{hyperref}
\usepackage{color}
\usepackage{amsmath}
\usepackage{graphicx}
\usepackage[english]{babel}
\usepackage{float}
\usepackage{url}
\usepackage{longtable}
\usepackage{fancyhdr}
\usepackage[normalem]{ulem}
\usepackage{comment}
\usepackage{tablefootnote}
\usepackage{lscape}

\usepackage{natbib,twoopt}
\usepackage{array}     
\newcolumntype{C}{>{$}c<{$}}
\bibpunct{(}{)}{;}{a}{}{,}             %% natbib format for A&A and ApJ
\makeatletter
  \newcommandtwoopt{\citeads}[3][][]{\href{http://adsabs.harvard.edu/abs/#3}%
    {\def\hyper@linkstart##1##2{}%
     \let\hyper@linkend\@empty\citealp[#1][#2]{#3}}}
  \newcommandtwoopt{\citepads}[3][][]{\href{http://adsabs.harvard.edu/abs/#3}%
    {\def\hyper@linkstart##1##2{}%
     \let\hyper@linkend\@empty\citep[#1][#2]{#3}}}
  \newcommandtwoopt{\citetads}[3][][]{\href{http://adsabs.harvard.edu/abs/#3}%
    {\def\hyper@linkstart##1##2{}%
     \let\hyper@linkend\@empty\citet[#1][#2]{#3}}}
  \newcommandtwoopt{\citeyearads}[3][][]%
    {\href{http://adsabs.harvard.edu/abs/#3}
    {\def\hyper@linkstart##1##2{}%
     \let\hyper@linkend\@empty\citeyear[#1][#2]{#3}}}
\makeatother

\makeatletter

\def\instrefs#1{{\def\scsep{\def\scsep{,}}\@for\w:=#1\do{\scsep\ref{inst:\w}}}}
\renewcommand{\inst}[1]{\unskip$^{\instrefs{#1}}$}

\let\orgautoref\autoref
\renewcommand{\autoref}
        {\def\equationautorefname{Eq.}%
         \def\figureautorefname{Fig.}%
         \def\sectionautorefname{Sect.}%
         \def\subsectionautorefname{Sect.}%
         \def\subsubsectionautorefname{Sect.}%
         \orgautoref}

\defcitealias{Madhusudhan2023ApJ...956L..13M}{M23}
\defcitealias{Madhusudhan2025arXiv250412267M}{M25}
\defcitealias{Schmidt2025arXiv250118477S}{S25}
\defcitealias{Welbanks2025}{Welbanks, Nixon et al. (2025)}
\defcitealias{Pica-Ciamarra2025arXiv250510539P}{P-C25}

\newcommand*\samethanks[1][\value{footnote}]{\footnotemark[#1]}

\makeatother
\begin{document}

\title{Insufficient evidence for DMS and DMDS in the atmosphere of K2-18\,b}
\subtitle{From a joint analysis of JWST NIRISS, NIRSpec, and MIRI observations}

\author{
R.~Luque\inst{uchicago}\thanks{NHFP Sagan Fellow; \email{rluque@uchicago.edu}} \and
C.~Piaulet-Ghorayeb\inst{uchicago,equal}\thanks{E.\ Margaret Burbridge Postdoctoral Fellow}  \and
M.~Radica\inst{uchicago,equal}\thanks{NSERC Postdoctoral Fellow}  \and
Q.~Xue\inst{uchicago,equal}  \and
M.~Zhang\inst{uchicago,equal}\thanks{51 Pegasi b Postdoctoral Fellow}  \and
J.\,L.~Bean\inst{uchicago}  \and
D.~Samra\inst{uchicago}  \and
M.\,E.~Steinrueck\inst{uchicago}\samethanks[4]
}

\institute{
\label{inst:uchicago}Department of Astronomy \& Astrophysics, University of Chicago, Chicago, IL 60637, USA \and 
\label{inst:equal}These authors have contributed equally.
}
\date{Received 19 May 2025 / Accepted DD MM YYYY}

\abstract
{Recent JWST observations of the temperate sub-Neptune K2-18\,b have been interpreted as suggestive of a liquid water ocean with possible biological activity. Signatures of dimethyl sulfide (DMS) and dimethyl disulfide (DMDS) have been claimed in the near-infrared (using the NIRISS and NIRSpec instruments) and mid-infrared (using MIRI). However, the statistical significance of the atmospheric imprints of these potential biomarkers has yet to be quantified from a joint analysis of the entire planet spectrum.}
{We test the robustness of the proposed DMS/DMDS detections by simultaneously modeling the NIRISS and NIRSpec observations jointly with the MIRI spectrum, considering different data reductions and modeling choices.}
{We use three well-tested pipelines to re-reduce the JWST observations, and two retrieval codes to analyze the resulting transmission spectra as well as previously published data.}
{Our joint analysis of the panchromatic (0.6 -- 12\,µm) spectrum of K2-18\,b finds insufficient evidence for the presence of DMS and/or DMDS in the atmosphere of the planet. Furthermore, other molecules containing methyl functional groups (e.g., ethane) with absorption bands similar to DMS/DMDS provide an equally good fit to the data. We find that any marginal preferences are the result of limiting the number of molecules considered in the model and oversensitivity to small changes between data reductions.}
{Our results confirm that there is no statistical significance for DMS or DMDS in K2-18\,b's atmosphere. While previous works have demonstrated this on MIRI or NIRISS/NIRSpec observations alone, our analysis of the full transmission spectrum does not support claims of potential biomarkers. Using the best-fitting model including DMS/DMDS on the published data, we estimate that $\sim$25 more MIRI transits would be needed for a 3$\sigma$ rejection of a flat line relative to DMS/DMDS features in the planet's mid-infrared transmission spectrum. }


\maketitle
\nolinenumbers
%%%%%%%%%%%%%% INTRODUCTION %%%%%%%%%%%%%%
\section{Introduction} \label{sec:intro}

\begin{figure*}[ht!]
    \centering
    \includegraphics[width=0.99\hsize]{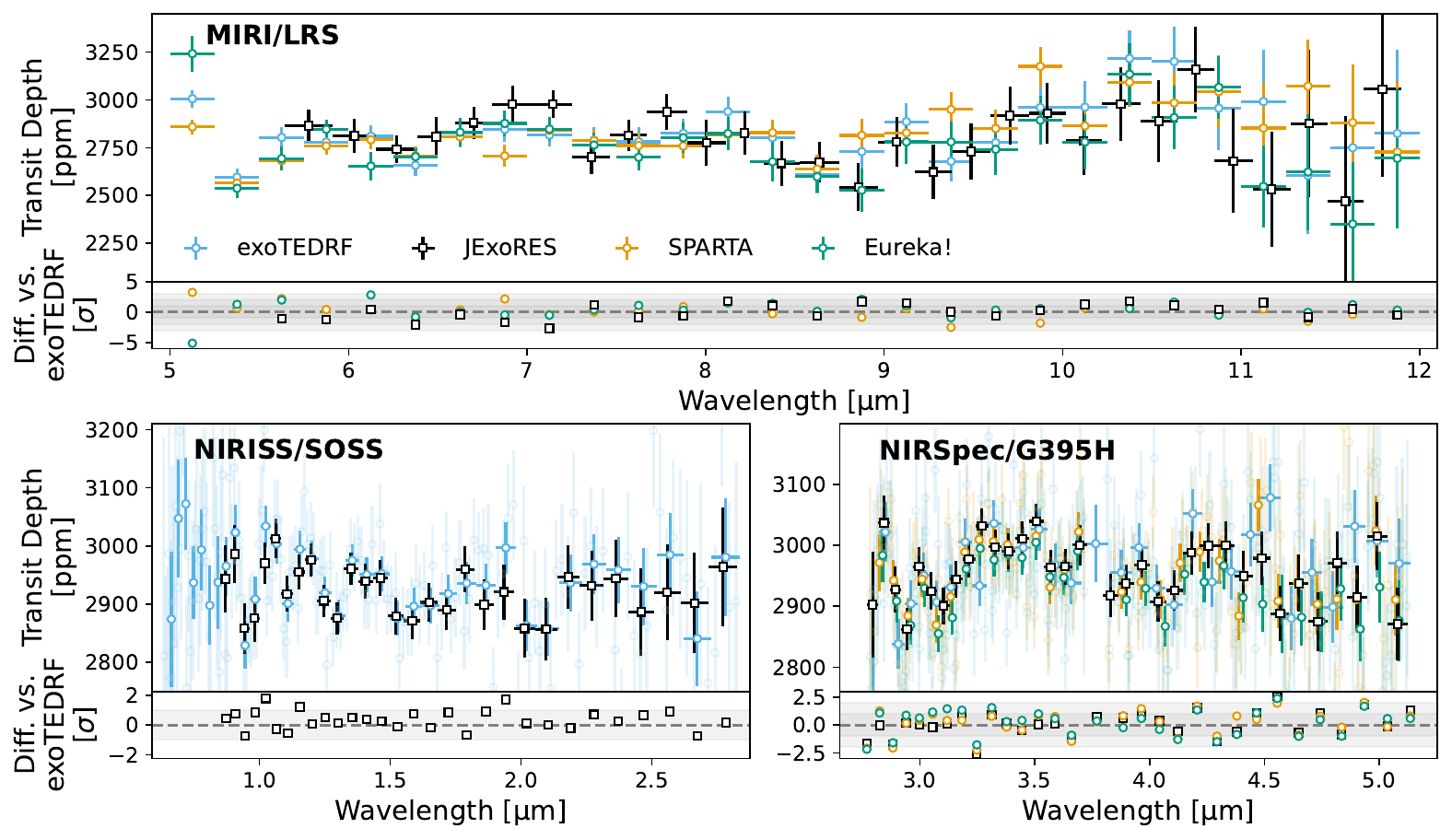}
    \caption{Transmission spectra of K2-18\,b from \texttt{exoTEDRF} (blue), \texttt{JExoRES} (black), \texttt{SPARTA} (orange) and \texttt{Eureka!} (green). For NIRISS and NIRSpec, faded points show higher resolution ($R$=100 and 200, respectively) spectra, and solid colors are binned for clarity. Smaller panels show error-normalized differences relative to \texttt{exoTEDRF}. The \texttt{JExoRES} spectra used different binning and are interpolated in the difference panels.    }
    \label{fig:transmission_spectrum}
\end{figure*}

The temperate\footnote{$T_{\rm eq} = 255\pm4\,\mathrm{K}$ assuming 0.3 Bond albedo \citep{Benneke2019ApJ...887L..14B}.} sub-Neptune K2-18\,b \citep{Montet2015ApJ...809...25M} %,Benneke2017ApJ...834..187B,Cloutier2017A&A...608A..35C,Sarkis2018AJ....155..257S,Cloutier2019A&A...621A..49C,Radica2022MNRAS.517.5050R}
continues to be one of the most intensively studied exoplanets in the search for habitable environments beyond the solar system. Initial Hubble Space Telescope transmission spectroscopy suggested the presence of water vapor in the planet's atmosphere \citep{Benneke2019ApJ...887L..14B, tsiaras_water_2019}, though later analyses highlighted degeneracies between H$_2$O and CH$_4$ \citep{barclay_stellar_2021, Blain2021}. With the increased wavelength range of JWST, \citet[][hereafter M23]{Madhusudhan2023ApJ...956L..13M} reported the detection of CH$_4$ at 5$\sigma$ and CO$_2$ at 3$\sigma$ in the H$_2$-rich atmosphere of K2-18\,b using data between 0.8 and 5\,µm. These findings were interpreted as evidence supporting a hycean scenario --- a world covered by a water ocean beneath a thin H$_2$-rich atmosphere --- potentially offering habitable conditions \citep{PietteMadhusudhan2020,madhusudhan_habitability_2021}. 

\citetalias{Madhusudhan2023ApJ...956L..13M} also reported a low-significance signal from dimethyl sulfide (C$_2$H$_6$S or DMS) --- a potential biosignature gas --- in their near-infrared (NIR) spectrum. However, the robustness of these detections, particularly the tentative inference of DMS, has been the subject of active scrutiny. A reanalysis of the data presented in \citetalias{Madhusudhan2023ApJ...956L..13M} by \citet[][hereafter S25]{Schmidt2025arXiv250118477S} confirmed CH$_4$ at 4$\sigma$, but found no statistically significant evidence for either CO$_2$ or DMS. Even using the molecular abundances from \citetalias{Madhusudhan2023ApJ...956L..13M}, the hycean interpretation itself has been challenged \citep{wogan_jwst_2024,Shorttle2024,huang_probing_2024}.

More recently, \citet[][hereafter M25]{Madhusudhan2025arXiv250412267M} inferred spectral features consistent with DMS and/or dimethyl disulfide (C$_2$H$_6$S$_2$ or DMDS) at 3$\sigma$ confidence in new mid-infrared JWST data from 5 to 12\,µm. \citetalias{Madhusudhan2025arXiv250412267M} argued that they represent independent evidence for the presence of sulfur-bearing biosignature gases. However, this analysis did not include the existing NIR JWST data, and the statistical robustness of the detection itself has been brought into question by \citet{taylor_2025} and \citetalias{Welbanks2025}. In this work, we present a reanalysis of the full suite of JWST transmission spectra of K2-18\,b to reassess the evidence for DMS and/or DMDS. Advanced data products from this paper can be downloaded from \href{https://doi.org/10.5281/zenodo.15304824}{Zenodo}.

%%%%%%%%%%%%%% FIGURES %%%%%%%%%%%%%%

\begin{figure*}[ht!]
    \centering
    \includegraphics[width=0.91\hsize]{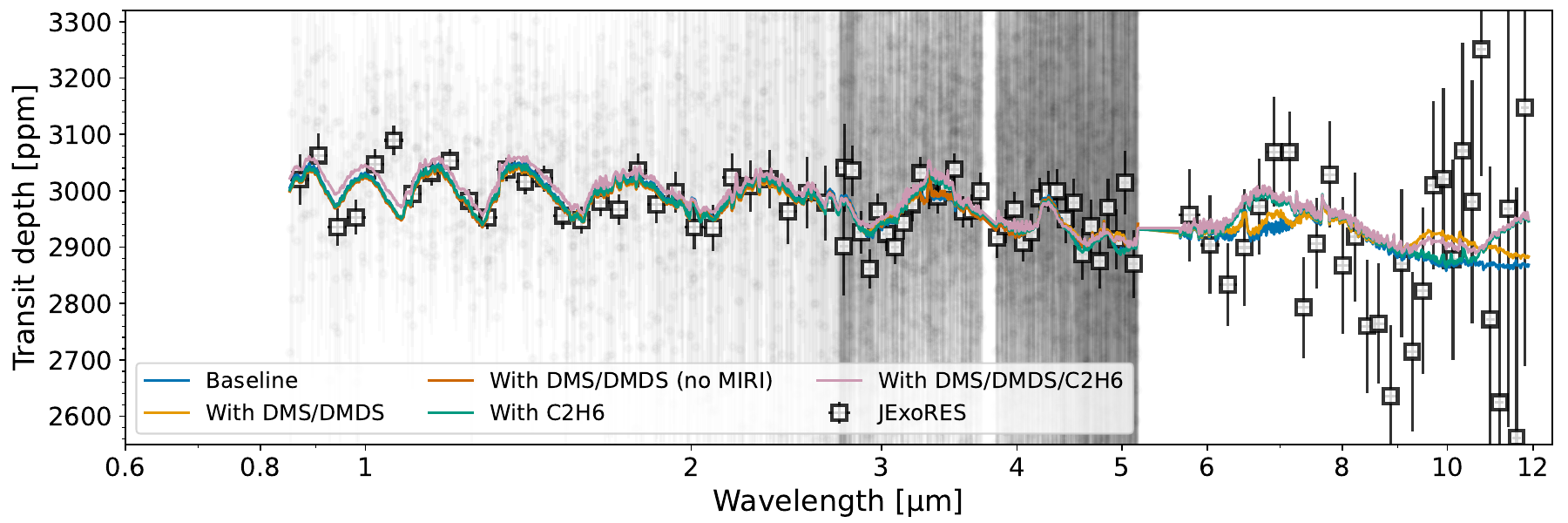}
    \includegraphics[width=0.91\hsize]{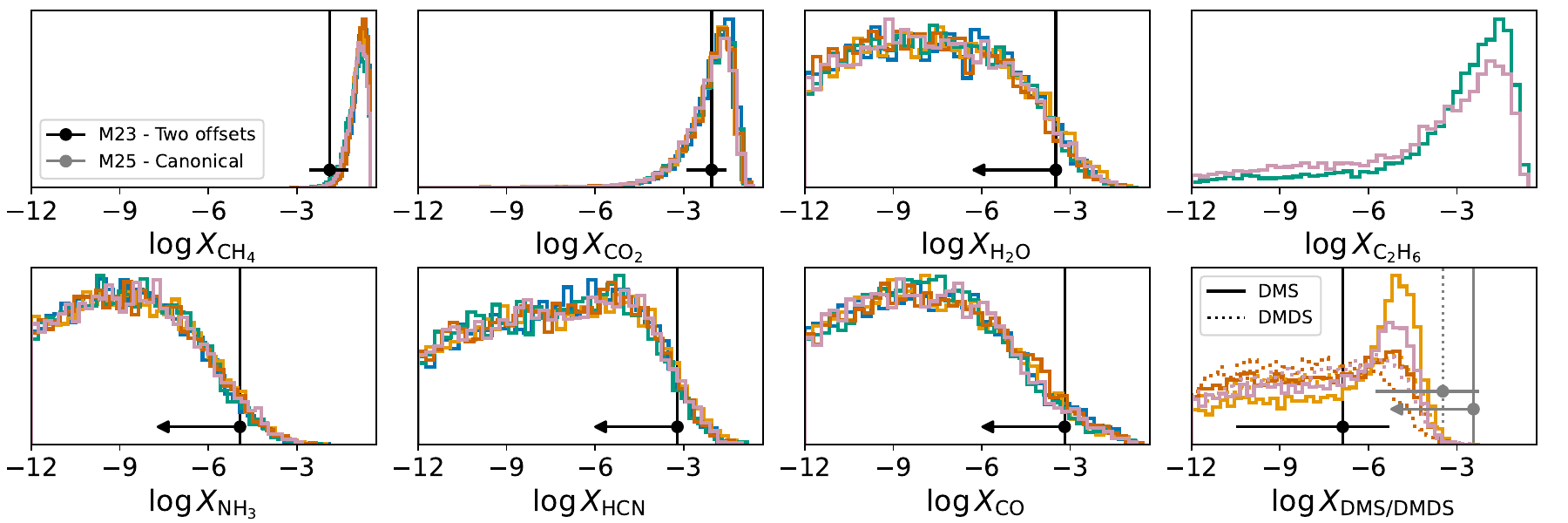}
    \caption{Results from the PLATON retrievals performed on the transmission spectrum of K2-18\,b produced with \texttt{JExoRES} in \citetalias{Madhusudhan2023ApJ...956L..13M} and \citetalias{Madhusudhan2025arXiv250412267M}. \textit{Top:} Transmission spectrum (native-resolution data used in the retrievals in gray, binned data in black for visualization purposes). Best-fit models for the five cases outlined in the text are shown: a baseline model (blue), a model with DMS and DMDS (orange), a model with a common hydrocarbon C$_2$H$_6$ (green), a model with DMS and DMDS but without using the MIRI data (red), and the baseline model with DMS, DMDS, and C$_2$H$_6$ (pink). Models are smoothed to $R\sim200$ for clarity. \textit{Bottom:} 1D marginalized posterior distributions on the molecular species included in each retrieval. Colors reflect the model in the top panel. The black points and vertical dashed lines indicate the abundances reported for each molecule in \citetalias{Madhusudhan2023ApJ...956L..13M} from just NIRISS and NIRSpec, whereas the gray points in the DMS/DMDS panel indicate the abundances reported in \citetalias{Madhusudhan2025arXiv250412267M} from MIRI data alone.}
    \label{fig:platon_retrievals}
\end{figure*}

\begin{figure*}[ht!]
    \centering
    \includegraphics[width=0.91\hsize]{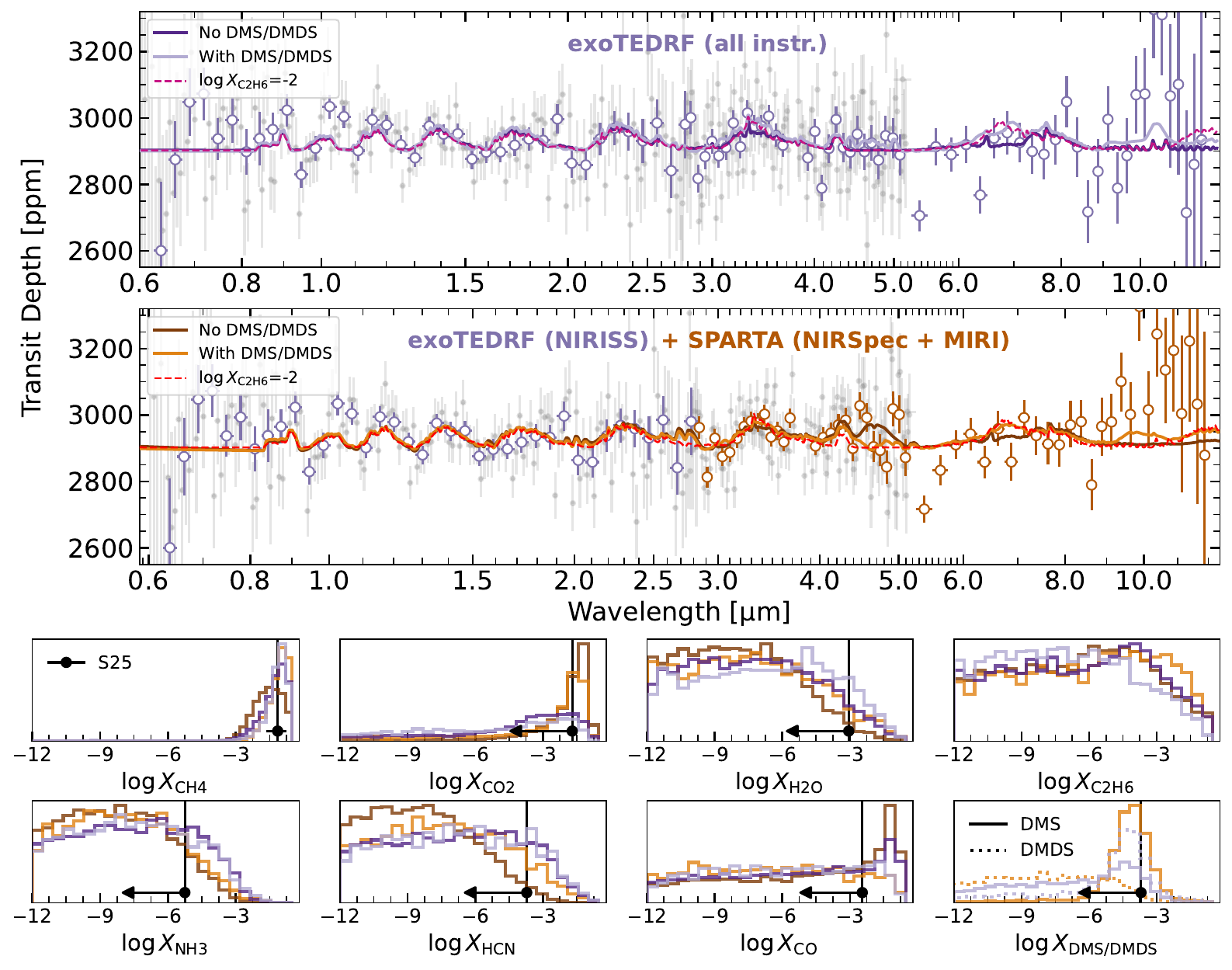}
    \caption{Results from the SCARLET retrievals on  \texttt{exoTEDRF} and \texttt{SPARTA} where C$_2$H$_6$ is included in the baseline. \textit{Top:} Retrievals performed on the \texttt{exoTEDRF} dataset (data in gray, binned in purple). The best-fit model for the retrievals without (with) DMS or DMDS is shown in dark (light) purple, smoothed to $R=200$. We show as a dashed line the impact of increasing the C$_2$H$_6$ abundance to $10^{-2}$ in the best-fit model from the retrieval without DMS/DMDS with the dashed line. \textit{Middle:} Same, but using \texttt{SPARTA} (orange) for the NIRSpec and MIRI datasets. \textit{Bottom:} 1D marginalized posterior distributions on the molecular species included in each retrieval. Colors reflect the models in the top. Black error bars and vertical lines (median $\pm1\sigma$ or 2$\sigma$ upper limits) correspond to the combined \texttt{POSEIDON} results from \citetalias{Schmidt2025arXiv250118477S} (which did not include the MIRI data). }
    \label{fig:scarlet_retrievals}
\end{figure*}

%%%%%%%%%%%%%% JWST OBSERVATIONS %%%%%%%%%%%%%%
\section{Data} \label{sec:data}

The observations of K2-18\,b analyzed in this work \citepalias[and in][]{Madhusudhan2023ApJ...956L..13M,Madhusudhan2025arXiv250412267M,Schmidt2025arXiv250118477S} are part of JWST GO Program 2722. There are a total of three transit observations: one each with the Near Infrared Imager and Slitless Spectrograph \citep[NIRISS;][]{Doyon2023} in Single Object Slitless Spectroscopy \citep[SOSS;][]{Albert2023} mode, the Near Infrared Spectrograph \citep[NIRSpec;][]{Jakobsen2022} with the G395H grating, and the Mid-Infrared Instrument \citep[MIRI;][]{Wright2023} in Low Resolution Spectroscopy \citep[LRS;][]{Bouwman2023} mode, which yields a continuous 0.6--12\,µm spectrum. For more information about the specific observing setups, see \citetalias{Madhusudhan2023ApJ...956L..13M} and \citetalias{Madhusudhan2025arXiv250412267M}. In this work, we analyze the \texttt{JExoRES} pipeline results presented in \citetalias{Madhusudhan2023ApJ...956L..13M} and \citetalias{Madhusudhan2025arXiv250412267M} and produce independent transmission spectra using \texttt{exoTEDRF} \citep{exoTEDRF} for NIRISS, NIRSpec, and MIRI\footnote{\texttt{exoTEDRF} v2.3.0 adds support for MIRI, making it the only publicly available pipeline to fully support NIRISS, NIRSpec, and MIRI.}; as well as \texttt{Eureka!} \citep{Eureka} and \texttt{SPARTA} \citep{Kempton2023Natur.620...67K} for NIRSpec and MIRI. Our transmission spectra are shown in Fig.~\ref{fig:transmission_spectrum}. % and available in \href{https://doi.org/10.5281/zenodo.15304824}{Zenodo}. 
In Appendix~\ref{app:reduction}, we summarize relevant choices in each reduction and light curve modeling from each pipeline.

%%%%%%%%%%%%%% ANALYSIS %%%%%%%%%%%%%%
\vspace{-5mm}
\section{Analysis} \label{sec:retrievals}

We carry out two main modeling analyses. The first is on the \texttt{JExoRES} native-resolution NIRISS and NIRSpec transmission spectra from \citetalias{Madhusudhan2023ApJ...956L..13M} and the MIRI transmission spectrum from \citetalias{Madhusudhan2025arXiv250412267M}. For this, we use PLATON (see Appendix~\ref{subsec:setup_platon}) %\citep{zhang_2019,zhang_2020,zhang_retrievals_2024} 
because it is fast enough to run retrievals at the high spectral resolution required. The second is on our independent data reductions with the SCARLET retrieval framework (see Appendix~\ref{subsec:setup_scarlet}). %The retrieval outputs and corner plots with posteriors for all parameters of these analyses are available in \href{https://doi.org/10.5281/zenodo.15304824}{Zenodo}. 

We follow \citet{2008ConPh..49...71T} to assess the statistical significance of our findings using Bayesian model comparison of log-evidences. Defining $\Delta \ln Z \equiv \ln Z_{\rm M1} - \ln Z_{\rm M0}$, we use the following thresholds: $\Delta \ln Z < 1$, statistically indistinguishable; $\Delta\ln Z \in [1,2.5)$ weakly favored; $\Delta\ln Z \in [2.5,5)$, moderately; and $\Delta\ln Z \geq 5$, strongly favored. We highlight that direct representation of these values to a sigma-value to establish the strength of any detection would be misleading \citep[e.g.,][]{benneke_how_2013,Welbanks2023AJ....165..112W}.

\vspace{-5mm}
\paragraph{PLATON.} \label{subsec:platon_retrievals}
Starting with a baseline model including CH$_4$, CO$_2$, H$_2$O, NH$_3$, HCN, and CO, we perform five isothermal free retrievals: on the baseline model, one including DMS and DMDS, one with both but excluding MIRI, one without DMS/DMDS but with C$_2$H$_6$ (ethane), and another with all three. The retrieval setup is described in Appendix~\ref{subsec:setup_platon}, while the best-fit models and the abundance posteriors for each molecule are shown in Fig.~\ref{fig:platon_retrievals} and Table~\ref{table:abundances}. While all the PLATON retrievals presented here model the impact of the transit-light-source effect (TLSe; e.g., \citealp{rackham_transit_2018}) on the spectrum, an identical retrieval was performed without TLSe, which yielded indistinguishable molecular abundance posteriors, in line with the findings of \citetalias{Schmidt2025arXiv250118477S}. % and the absence of correlation between TLSe parameters and abundances in the corner plots provided in \href{https://doi.org/10.5281/zenodo.15304824}{Zenodo}. 

\vspace{-5mm}
\paragraph{SCARLET.} \label{subsec:scarlet_retrievals}
Given the similarities between the transmission spectra derived in this work (Fig.~\ref{fig:transmission_spectrum}), we use only two combinations for this analysis: one using \texttt{exoTEDRF} for all instruments; and another using \texttt{exoTEDRF} for NIRISS and \texttt{SPARTA} for NIRSpec and MIRI. We perform four isothermal free retrievals on each transmission spectrum using the same baseline model and combinations as before (except for the `no MIRI' case). The retrieval setup is described in Appendix~\ref{subsec:setup_scarlet}, and the best-fit models and abundance posteriors are shown in Fig.~\ref{fig:scarlet_retrievals} and Table~\ref{table:abundances}.

%%%%%%%%%%%%%% RESULTS AND DISCUSSION %%%%%%%%%%%%%%
\section{Results and Discussion} \label{sec:discussion}

% Header removed for space reasons
%\subsection{Impact of data reduction choices} \label{subsec:pipelines}

Our retrievals of the \texttt{JExoRES} reduction of the 0.8--12\,µm transmission spectrum disfavor the presence of DMS/DMDS. Including them relative to a model without those species causes $\ln Z$ to decrease by 0.3, indicating no evidence for these species. Data reduction choices have a moderate impact on the qualitative conclusions. When considering the \texttt{exoTEDRF} and \texttt{exoTEDRF+SPARTA} spectra, the inclusion of DMS/DMDS results in a positive Bayes factor (i.e., $\Delta \ln Z = 2.3$--2.8). Small changes across data reductions at 7\,µm (DMS preferred for \texttt{exoTEDRF+SPARTA}) and at 10\,µm (DMDS preferred for \texttt{exoTEDRF}-only) are likely responsible for the discrepancy. In no case does the preference for DMS/DMDS reach $\Delta \ln Z>5$, considered standard for any molecule detection claim. 

The baseline model did not include the many other hydrocarbons that are expected chemically in the atmosphere. Thus, our values are likely inflated due to a lack of inclusion of other candidate molecules such as propyne and diethyl sulfide, which have been proposed to explain the MIRI \citepalias{Welbanks2025} and NIRISS+NIRSpec \citep[][hereafter P-C25]{Pica-Ciamarra2025arXiv250510539P} spectra, respectively. We consider one additional candidate (ethane) below, as an exhaustive search for all molecules in the full spectrum is beyond the scope of this work.

All reductions (\texttt{JExoRES}, \texttt{exoTEDRF}, \texttt{exoTEDRF+SPARTA}) prefer a high CH$_4$ abundance, but show no evidence of H$_2$O, NH$_3$, HCN, or CO (Figs.~\ref{fig:platon_retrievals} and \ref{fig:scarlet_retrievals}). The  \texttt{SPARTA} and \texttt{JExoRES} versions of the spectra yield two-sided posteriors on the CO$_2$ abundance, whereas \texttt{exoTEDRF} yields broader CO$_2$ posteriors. The inferred volume mixing ratios (VMR) in our retrievals on the same data are consistent with those from the ``two-offset'' case in \citetalias{Madhusudhan2023ApJ...956L..13M} (CH$_4$ median higher by 1\,dex, CO$_2$ by 0.1\,dex). 

Without DMS/DMDS, we find a slight posterior peak at high CO abundances in our data reductions, but still consistent with a non-detection. Since this same peak does not arise in the PLATON retrievals, it may be, instead, an artifact of the model resolution ($R=20,000$ was used for SCARLET vs.\ $R=100,000$ for PLATON), as CO is better constrained using the native resolution NIRSpec spectrum \citep[e.g.,][]{esparzaborges_CO_2023}. We highlight that a peak in a posterior does not mean the detection of a chemical absorber \citep[e.g.,][]{Welbanks2022}.

\subsection{Uniqueness of DMS/DMDS absorption features} \label{subsec:hydrocarbons}

Ethane (C$_2$H$_6$) is a common byproduct of methane photolysis by recombining two methyl radicals, and the most abundant hydrocarbon predicted by the abiotic photochemistry simulations of K2-18\,b in \citet{huang_probing_2024}; \citetalias{Welbanks2025}. Ethane has a very similar chemical structure to DMS. The latter consists of two methyl groups with a sulfur atom in the middle, while the former consists of two methyl groups connected directly. The similarity in molecular structure gives rise to similarities in their absorption cross sections: the locations of the strongest absorption band in DMS (a C-H stretching mode at 3.4\,µm) as well as the second strongest (a H-C-H bending mode at 6.9\,µm) are both attributable to the methyl functional group \citep[Table 1.2;][]{tammer_2004}. Therefore, they are present for DMS, DMDS, C$_2$H$_6$ (Fig. \ref{fig:scarlet_retrievals}), and large classes of other organic molecules \citep{sousa_silva_molecular_2019}. 

Although the models with or without ethane are statistically indistinguishable, we even find a slight model preference for the inclusion of ethane in retrievals performed on the \texttt{JExoREs} and \texttt{exoTEDRF} datasets (Table \ref{table:abundances}). 
% Figure~\ref{fig:scarlet_retrievals} also shows how a high ethane abundance can mimic a DMS or DMDS detection. %This preference is driven by the slightly better $\chi^2 / N=1.0757$ for the model including ethane compared to $\chi^2 / N =1.0766$ for the baseline model, and $\chi^2 / N = 1.0759$ when adding DMS/DMDS. 
Our results strengthen the conclusion that a wide range of photochemically plausible hydrocarbons could reasonably explain the observations, in line with the analyses from \citetalias{Welbanks2025} and \citetalias{Pica-Ciamarra2025arXiv250510539P}.

\subsection{Impact of the addition of MIRI/LRS} \label{subsec:miri}
As shown in Fig.~\ref{fig:platon_retrievals}, the addition of MIRI data has a negligible impact on the posteriors of all molecules except DMS and DMDS, compared to those from the NIRISS and NIRSpec data alone. %The bottom panels of Fig.~\ref{fig:platon_retrievals} show the reported abundances in \citetalias{Madhusudhan2023ApJ...956L..13M} and \citetalias{Madhusudhan2025arXiv250412267M} for comparison. 
In the \texttt{JExoRES} retrievals, the DMS posterior become more sharply peaked around VMR $\sim 10^{-5}$ with the addition of MIRI, but remains an upper limit. %as in in all of our independent reductions (Fig.~\ref{fig:scarlet_retrievals}).

Furthermore, we retrieve photospheric temperatures much lower than the $T_p = 422_{-133}^{+141}$\,K at 1\,mbar inferred by \citetalias{Madhusudhan2025arXiv250412267M} from the MIRI data alone --- which is inconsistent with the predicted zero-albedo, no-atmosphere substellar temperature of $T_{\rm eff}\sqrt{R_\star/a}=380$\,K. This conclusion is supported regardless of the adopted photosphere temperature prior adopted\footnote{See Appendix~\ref{subsec:setup_platon} for a discussion of the temperature priors and posterior constraints.}, and across PLATON and SCARLET retrievals. This discrepancy is driven by the fact that the apparent ``features" in the MIRI spectrum are much larger than the spectral features at bluer wavelengths. If they were indeed from the planet, the atmosphere would need to be relatively hot to have a sufficiently large scale height. However, this is inconsistent with both energy balance and the more precise NIRISS and NIRSpec observations. Therefore, while the MIRI features are suggestive of DMS, DMDS, or other hydrocarbons, and thus increase the evidence for these species in the retrievals, the features are implausibly large.

We performed a suite of simulations to determine how many MIRI transits would be needed to robustly detect spectral features, assuming the best-fitting model from the PLATON DMS/DMDS retrieval on the \texttt{JExoRES} spectrum represents the underlying atmosphere. We find that an average of 26 transits have to be stacked to obtain a 3$\sigma$ rejection of a flat line when scaling the \texttt{JExoRES} errors. However, this result is highly sensitive to the noise realization, and the standard deviation of the number of transits needed in 1,000 simulations is 13. We note that due to the star's position in the sky and the long orbital period of the planet, there are typically only four transit opportunities observable with JWST each year.

%%%%%%%%%%%%%% CONCLUSIONS %%%%%%%%%%%%%%
\section{Conclusions} \label{sec:conclusions}

K2-18\,b is undoubtedly one of the most promising planets for interior, atmosphere, and habitability studies of a sub-Neptune in the JWST era. Transmission spectroscopy robustly finds CH$_4$ in its hydrogen-rich atmosphere, but the presence of other species is still under debate. In this letter, we carried out an independent analysis at both the data reduction and atmospheric modeling level, leveraging the full wavelength coverage (0.6--12\,µm) publicly available for this planet. We can reproduce the abundances reported by \citetalias{Madhusudhan2023ApJ...956L..13M} for CH$_4$ and CO$_2$ using either the previous spectrum or our independent reductions, although the exact abundance of CO$_2$ is influenced by the specific pipeline used. DMS/DMDS are not detected robustly across different data reductions or molecules included in the baseline model, consistent with the results from \citetalias{Pica-Ciamarra2025arXiv250510539P} for NIRISS/NIRSpec when adding instrumental offsets and \citetalias{Welbanks2025} for MIRI.

%independent analyses of the MIRI data by \citet{taylor_2025} and \citetalias{Welbanks2025}.

We find that other generic hydrocarbons can provide an alternative explanation for the putative DMS/DMDS spectral features proposed by \citetalias{Madhusudhan2023ApJ...956L..13M} and \citetalias{Madhusudhan2025arXiv250412267M}, as demonstrated with the case of ethane (C$_2$H$_6$) in this work. Further, no DMS/DMDS cross-sections are yet available for pressures lower than 1\,bar atmosphere which would be more representative of the upper atmosphere probed in transmission, leading to overestimated DMS/DMDS absorption features. Oppositely, the line lists for most organic molecules and hydrocarbons, including ethane, are extremely incomplete to date, which can lead to an underestimation of their absorption signal. The clear deficiencies in the current line lists highlight the need for further experimental and theoretical studies to address them, particularly as next-generation facilities like HWO and the ELTs aim to supercharge the search for habitable conditions on other planets.

%%%%%%%%%%%%%% ACKNOWLEDGEMENTS %%%%%%%%%%%%%%
\begin{acknowledgements}
The authors would like to thank Guangwei Fu, Matt Nixon, Brandon Park Coy, Vivien Parmentier, and Luis Welbanks for helpful discussions. This work is based on observations made with the NASA/ESA/CSA James Webb Space Telescope. The data were obtained from the Mikulski Archive for Space Telescopes at the Space Telescope Science Institute, which is operated by the Association of Universities for Research in Astronomy, Inc., under NASA contract NAS 5-03127 for JWST. R.L. is supported by NASA through the NASA Hubble Fellowship grant HST-HF2-51559.001-A awarded by the Space Telescope Science Institute, which is operated by the Association of Universities for Research in Astronomy, Inc., for NASA, under contract NAS5-26555. M.Z. and M.E.S. thank the Heising-Simons Foundation for support through the 51 Pegasi b fellowship. C.P.-G. is supported by the E. Margaret Burbidge Prize Postdoctoral Fellowship from the Brinson Foundation. M.R.\ acknowledges financial support from the Natural Sciences and Engineering Research Council of Canada through a Postdoctoral Fellowship. He would also like to thank Hannah Wakeford and David Grant for useful conversations about MIRI systematics. 
\end{acknowledgements}

%\clearpage\newpage
\vspace{-5mm}
%%%%%%%%%%%%%% REFERENCES %%%%%%%%%%%%%%
\bibliographystyle{aa} % style aa.bst
\bibliography{biblio} % your references biblio.bib

%%%%%%%%%%%%%% APPENDICES %%%%%%%%%%%%%%

\begin{appendix} %First online appendix

\section{Reduction Details}
\label{app:reduction}

\subsection{\texttt{JExoRES}}

We use the NIRISS/SOSS and NIRSpec/G395H spectra presented in \citetalias{Madhusudhan2023ApJ...956L..13M} using \texttt{JExoRES} \citep{JExoRES} that can be downloaded \href{https://osf.io/36djh/files/osfstorage}{here}. For the MIRI/LRS spectra, we use the \texttt{JExoRES} version presented in \citetalias{Madhusudhan2025arXiv250412267M} that can be downloaded \href{https://osf.io/gmhw3/files/osfstorage}{here}. The \texttt{JexoPipe} pipeline \citep{JexoPipe} was also used in \citetalias{Madhusudhan2025arXiv250412267M}, but these results are not public. 

\subsection{\texttt{exoTEDRF}}

As of release 2.3.0, \texttt{exoTEDRF} \citep{exoTEDRF, radica_awesome_2023, feinstein_early_2023} is now the first publicly available pipeline for exoplanet time series observations with full support for NIRISS, NIRSpec, and MIRI. We thus make use of \texttt{exoTEDRF} to produce a 0.6 -- 12\,µm spectrum with a single data analysis framework. 

\subsubsection{NIRISS/SOSS \& NIRSpec/G395H.}
We use the NIRISS/SOSS and NIRSpec/G395H \texttt{exoTEDRF} spectra from \citetalias{Schmidt2025arXiv250118477S}. For details on the data reduction and light curve analysis, we point the reader to Sections 2.2, 3.2, and Table~1 of that work. Here, we use the transmission spectrum computed at a resolution of $R=100$ for NIRISS and $R=200$ for NIRSpec. This choice is motivated by the analysis in \citetalias{Schmidt2025arXiv250118477S} (their Fig.~5) showing that this is the configuration that results in the highest detection significances of CH$_4$, CO$_2$, and DMS.

% made publicly available\footnote{\url{https://doi.org/10.5281/zenodo.14735688}}

\subsubsection{MIRI/LRS.}
As this is the first use of \texttt{exoTEDRF} for MIRI observations, we briefly describe the methodology below, and refer the reader to Radica et al.~(2025, in prep.) for more details.

In \texttt{Stage 1}, we perform data quality and saturation flagging, EMI correction, linearity correction, dark current subtraction, cosmic ray flagging, and ramp fitting. During linearity correction, we flag the first five and last groups up-the-ramp as \texttt{DO\_NOT\_USE} before proceeding with the linearity correction using the default STScI polynomials. As in \citetalias{Madhusudhan2025arXiv250412267M}, we find that applying a linearity self-calibration results in a consistent spectrum, as does varying the number of groups clipped at the start of the ramp. Cosmic ray identification is performed using a time-domain rejection algorithm, and a 7$\sigma$ threshold \citep{radica_muted_2024}. All other steps are the \texttt{exoTEDRF} defaults. We skip the gain scale correction since absolute flux values are not necessary for this work.

In \texttt{Stage 2}, we assign the WCS and source type, correct flat fielding effects, and interpolate bad pixels. We subtract the background using a row-wise median, calculated from two 14-pixel-wide regions on either side of the trace (columns 12--26 and 46--60). Finally, we perform a PCA reconstruction on the combined TSO data cube. This is similar to analyses performed by \citet{coulombe_broadband_2023} and \citet{radica_muted_2024}, except instead of using the PCA eigenvalue timeseries as vectors against which to detrend during light curve fitting, we reconstruct the 3D time series  (i.e., integration number, X-pixel, Y-pixel), removing components which are correlated with detector trends. For this particular timeseries, we remove the second component, which is correlated with a sub-pixel drift along the detector X (i.e., spatial) axis. Finally, we trace the spectrum on the detector using the \texttt{edgetrigger} algorithm \citep{radica_applesoss_2022}, and perform a box aperture extraction with a width of six pixels, clipping 4$\sigma$ outliers from the resulting light curves. 

We fit the light curves using the \texttt{exoUPRF} library \citep{RadicaexoUPRF, radica_promise_2025}. The white light curve is constructed using wavelengths 5--12\,µm. For the astrophysical transit model, we use a standard \texttt{batman} model \citep{batman}. We fix the period to 32.940045\,d \citep{Benneke2019ApJ...887L..14B}, and put Gaussian priors on the orbital parameters based on the best fitting values in \citetalias{Madhusudhan2023ApJ...956L..13M}. The limb darkening is freely fit in the range $u_1, u_2 \in [-1, 1]$ using the quadratic law. We include a systematics model consisting of an exponential ramp and linear slope, as well as an error inflation term added in quadrature to the extracted flux errors. 

For the spectroscopic fits, we bin the light curves in increments of 0.25\,µm ($R\sim100$), yielding 28 bins across the 5--12\,µm range. We fix the orbital parameters to the best-fitting values from the white light fits, and allow the scaled planet radius, transit zero point, error inflation, and two of the three parameters of the systematics model (linear slope and amplitude of exponential ramp) to vary freely. We, however, fix the exponential ramp timescale to the best-fitting values from the white light fit, as it remains unconstrained otherwise. We fix the two parameters of the quadratic limb darkening law to predictions from \texttt{ExoTiC-LD} \citep{david_grant_2022_7437681} using MPS-ATLAS2 stellar models \citep{Kostogryz2022}. We also cut the first 250 integrations from each spectroscopic light curve as in \citetalias{Madhusudhan2025arXiv250412267M}.

\subsection{\texttt{Eureka!}}

\subsubsection{NIRSpec/G395H.}
We reduce the observations and fit the spectroscopic light curves using an identical setup as the ``Eureka! B'' analysis in \citetalias{Schmidt2025arXiv250118477S}. For this work, we compute a set of spectroscopic light curves at $R=200$ (not made publicly available) for direct comparison with the other reductions. The control and parameter files for Stages 4--6 are available for download from \href{https://doi.org/10.5281/zenodo.15304824}{Zenodo}.

\subsubsection{MIRI/LRS.}
We start with the uncalibrated files and run Stages 1--2, which are wrappers of the STScI \texttt{jwst} pipeline. We use default steps (data quality initialization, electromagnetic interference correction, saturation flagging, discard first and last frames, linearity correction, reset switch charge decay correction, dark current subtraction, and ramp fitting) and a 15$\sigma$ threshold in the jump step. We do not perform group-level background subtraction. We skip the flat-field correction in Stage 2. We then extract the stellar time-series spectra in Stage 3. We perform a mean column-by-column background subtraction, excluding the area within 10 pixels of the center of the trace. We use an aperture of 9 pixels, corresponding to a 4-pixel half width, to optimally extract \citep{Horne1986} the time-series spectra. In Stage 4, we generate broadband (white light, covering 5.24--12.02\,µm) and spectroscopic light curves using a bin width of 0.25\,µm (between 5--12\,µm, equaling 28 spectroscopic bins). 

In Stage 5, we fit the white light curve using an exponential ramp and a linear trend as the systematics model. We mask the first 250 integrations to remove the strongest effect of the detector settling, assume a circular orbit, fix the period to 32.940043\,d \citep{Benneke2019ApJ...887L..14B}, and give uninformative priors to the rest of the parameters ($t_0$, $a/R_\star$, $i$, $R_p/R_\star$, $q_1$, and $q_2$). We use a quadratic limb darkening law parameterized as in \citet{Kipping13}. We add in quadrature an error inflation term to the extracted flux errors. Our choices mimic those of the final \texttt{JExoRES} reduction of \citetalias{Madhusudhan2025arXiv250412267M}. Our best-fit parameters are consistent within 1$\sigma$ with those of that work, except for $a/R_\star$ where we find $78.8\pm1.1$ as opposed to $80.3\pm0.5$ in \citetalias{Madhusudhan2025arXiv250412267M}. To obtain the transmission spectrum, we fix all orbital parameters (except $R_p/R_\star$ and the parameters associated with the systematics model) to the values from our white light curve analysis. We include the control and parameter files for Stages 1--6, as well as the white light curve and best-fit parameters of their model, in \href{https://doi.org/10.5281/zenodo.15304824}{Zenodo} to verify all our choices and reproduce these results.

\subsection{\texttt{SPARTA}}

\texttt{SPARTA} is an end-to-end data reduction and analysis pipeline for JWST. Starting from the \texttt{\_uncal.fits} files, \texttt{SPARTA} operates completely independently of the official JWST pipeline \citep{jwst_pipeline} and is 27 times faster than the \texttt{calwebb\_detector1} (JWST Stage 1) code. In this paper, we re-analyze both the NIRSpec and MIRI/LRS data using  \texttt{SPARTA}. 

\subsubsection{NIRSpec/G395H.}
We generate the \texttt{\_rateints.fits} files following the same procedure described in \citet{Kempton2023Natur.620...67K}, which includes applying the superbias (reference file version of 0479 for NRS1, 0474 for NRS2), correcting with the reference pixels, applying the non-linearity correction (0024), subtracting the dark current (0457/0470), masking bad pixels (0082/0080), multiplying by the gain (0025/0027), fitting the slope, subtracting the residuals, and re-fitting the slope. The only deviation from \citet{Kempton2023Natur.620...67K} is the addition of a group-level background subtraction step before slope fitting. To perform this, we first identified the Gaussian center of the 2D spectra via least-squares fitting. We then masked a region within seven pixels of the center and subtracted the median of the remaining background region from the entire image. 

After generating the \texttt{\_rateints.fits} files, we applied a second background subtraction using the same group-level strategy and measured the x- and y-positions of the spectral trace from the cleaned median frame. We then performed optimal extraction over columns 610--2044 of NRS1 and columns 4--2044 of NRS2, covering wavelength ranges of 2.749--3.717\,µm and 3.823--5.100\,µm, respectively. Following extraction, we interpolated over bad rows and removed bad rows and columns. As a final step in the reduction, we computed a detrended light curve by subtracting a median-filtered light curve (kernel size 50). Integrations deviating by more than 3$\sigma$ from zero in the detrended light curve were flagged as bad.

For light curve fitting, we fixed $t_0$, $P$, $a/R_\star$, and $i$ to the values listed in Table 1 of \citetalias{Madhusudhan2023ApJ...956L..13M}. We binned the light curves to a resolving power of $R \sim 200$, resulting in 123 spectroscopic channels. The first 30 integrations were discarded. We modeled systematics using $F_{\rm sys} = F_\star (1 + m(t - \bar{t}) + c_yy + c_xx)$, where we account for a linear trend with time and detector positional systematics. We treated the limb darkening coefficients as free parameters and fitted them using the formalism of \citet{Kipping13}.

\subsubsection{MIRI/LRS.}
We reduced the MIRI/LRS data following the same procedure described in the NIRSpec section, with two exceptions: MIRI does not require a superbias correction, and we did not apply group-level background subtraction. For a more detailed description of the \texttt{SPARTA} MIRI/LRS reduction, we refer readers to \citet{Kempton2023Natur.620...67K}, \citet{zhang_gj_2024}, and \citet{xue_jwst_2023}. The calibration reference files used were: non-linearity (version 0032), dark current (0102), mask (0036), and gain (0042). For integration-level background subtraction, we computed the median value of rows [10, 21] and [$-21$, $-10$], and subtracted this value from each integration.

Optimal extraction was performed within five pixels of the trace, from column 20 to 275. We applied the same spectral binning as in the \texttt{exoTEDRF} and \texttt{Eureka!} reductions. For the spectroscopic light curve fitting, we fixed $t_0$, $a/R_\star$, $i$, $u_1$, and $u_2$ to the best-fit values derived from the wavelength-integrated white light curve: 60426.12880\,d (BJD $-$ 2400000.5), 78.87, 89.53\,deg, 0.0352, and 0.1083, respectively. The first 250 integrations were excluded from the fitting. We adopted $F_{\rm sys} = F_\star (1 + Ae^{-(t - t_0)/\tau} + m(t - \bar{t}) + c_yy + c_xx)$ as the systematics model where we account for ramp-like behavior and detector positional variation, as it yielded the lowest RMS residuals among the various models we tested.

\section{Retrieval Setup}

\renewcommand{\arraystretch}{1.2}
\begin{table}[ht]
    \centering
    \caption{Fixed and free parameters in the PLATON and SCARLET retrievals.  We retrieve $\log{X}$ for CH$_4$, CO$_2$, H$_2$O, NH$_3$, HCN, and CO in the baseline retrievals. DMS, DMDS, and C$_2$H$_6$ are added to the model depending on the test. $F$, $U$, and $N$ indicate fixed, uniform, and normal prior distributions, respectively. The SCARLET retrievals did not fit for stellar contamination, which was previously not found to impact the atmospheric inferences for the NIRISS and NIRSpec spectrum of K2-18\,b \citep{Schmidt2025arXiv250118477S}.}
    \label{table:retrieval_priors}
    \begin{tabular}{l|C|C}
        \hline\hline
        Parameter & \text{PLATON priors} & \text{SCARLET priors} \\
        \hline
         $R_\star$ [$R_\odot$] & F(0.4445)   &   F(0.4445)  \\
         $T_{\rm eff}$ [K] & F(3457)   &    F(3457)  \\        
         $M_p$ [$M_\oplus$] & N(8.63, 1.35^2)   &   N(8.63, 1.35^2)  \\
         $R_p$ [$R_\oplus$] & U(2.4, 2.8)   &  \text{optimized}^a  \\
         $T_p$ [K] & U(200, 500)   &  U(200, 1000)   \\
         $\log P_{\rm cloud}$ [Pa] & U(0, 8)   &  U(-3, 6.98)   \\
         $\log{X}$ & U(-12, -0.522)   &   U(-12, -0.522)  \\
         \text{NIRISS offset [ppm]} & U(-200, 200)   &  U(-250, 250)   \\ 
         \text{NRS1 offset [ppm]} & U(-200, 200)   &  U(-250, 250)   \\ 
         \text{MIRI offset [ppm]} & U(-200, 200)   &  U(-250, 250)   \\    
         $T_{\rm spot}$ [K] & U(2000, 3457)   & \dots    \\
         $f_{\rm spot}$ & U(0, 0.2)   & \dots    \\
         \hline
    \end{tabular}
    \footnotesize{$^a$ SCARLET fits the planet radius for each combination of atmosphere model parameters, see Sec.~\ref{subsec:setup_scarlet}.}
\end{table}

\subsection{PLATON} \label{subsec:setup_platon}

PLATON is a GPU-accelerated atmospheric retrieval code \citep{zhang_2019,zhang_2020,zhang_retrievals_2024} that has been used in numerous works, including for JWST transmission spectra of mini-Neptunes (e.g., \citealt{davenport_2025}).  We use the latest version of the code (6.3.1), except that we significantly speed up the method that bins the model transit spectrum to the observed wavelength bins.

The default PLATON opacities are at a spectral resolution of 20,000, which is too low to retrieve on the native-resolution NIRISS and NIRSpec spectra.  They have a temperature spacing of 100\,K, uncomfortably coarse for a planet as cold as K2-18b.  We replace them with $R=100,000$ opacities generated from the DACE opacity database\footnote{\url{https://dace.unige.ch/opacityDatabase}}, and halve the temperature spacing to 50\,K.  For CH$_4$, we use the HITEMP 2020 line list \citep{HargreavesEtal2020apjsHitempCH4}, which (unlike the ExoMol line lists) covers the entire wavelength range spanned by the NIRISS data.  C$_2$H$_6$ is not in the DACE database, so we compute absorption cross sections with ExoCross from the HITRAN 2020 line list \citep{hitran_2020}.  DMS and DMDS have no published line lists, so we assume the absorption cross sections measured at 278\,K in a 1\,bar nitrogen atmosphere by \citet{sharpe_DMS_2004} are valid at all temperatures and pressures.  Because these cross sections are far more pressure broadened at 1\,bar than at K2-18\,b's photospheric pressure of $\sim$1\,mbar, we expect the retrieval to overestimate absorption and thus underestimate the DMS/DMDS abundance.  On the other hand, the C$_2$H$_6$ line list is highly incomplete, likely leading to underestimation of absorption and overestimation of its abundance.

We perform a free retrieval via nested sampling with an isothermal atmosphere, an opaque cloud deck, transit light source effect (TLSe) from starspots, and three instrumental offsets relative to NRS2 (NIRISS, NRS1, MIRI).  The TLSe is parameterized with a starspot temperature and a starspot covering fraction, with the non-spotted surface fixed at the $T_{\rm eff}$ of 3457\,K \citep{Cloutier2017A&A...608A..35C}.  Both the starspot spectrum and the non-starspot spectrum are obtained from the \texttt{BT-Settl} (AGSS2009) spectral grid that is included in PLATON.  All fixed and free parameters, together with the priors assumed for the free parameters, are listed in Table~\ref{table:retrieval_priors}.  The nested sampling was performed by \texttt{pymultinest} with 1000 live points \citep{pymultinest}.  We used the Bayesian evidence reported by Importance Nested Sampling, which can calculate the evidence with up to an order of magnitude higher accuracy \citep{feroz_2019}. 

In our analyses, we found that when temperatures down to 100\,K are allowed, the retrievals obtain a surprisingly low photosphere temperature, with medians and error bars within 5\,K of $175_{-40}^{+50}$\,K. However, the posteriors are sufficiently broad that the temperature is 2$\sigma$ consistent with the equilibrium temperature of $\sim$250\,K. To explore the effect of forcing higher temperatures, we perform a retrieval with DMS/DMDS where we impose a minimum temperature of 100\,K instead of 200\,K.  This slightly lowers the upper limits on DMS and DMDS, by 0.35 and 0.57 dex, respectively.  It also decreases the median CH$_4$ abundance by 0.23 dex, making it more similar to the SCARLET results.

\subsection{SCARLET} \label{subsec:setup_scarlet}

We analyze our independently derived transmission spectra with the SCARLET atmospheric retrieval framework \citep{benneke_atmospheric_2012,benneke_how_2013,benneke_strict_2015, Benneke2019ApJ...887L..14B,roy_water_2023-1} including modifications for more flexible modeling of sub-Neptune atmospheres \citep{piaulet_evidence_2023,piaulet-ghorayeb_jwstniriss_2024}. We elect to perform free retrievals, where we fit for well-mixed abundances of the gaseous species making up the atmosphere. This approach does not require that the relative abundances match chemical equilibrium, which might not be a valid assumption in the upper atmospheres of colder sub-Neptunes \citep{wogan_jwst_2024,benneke_jwst_2024}.

Our forward model iterates over the radiative transfer and hydrostatic equilibrium calculation to determine the altitude-pressure mapping consistent with the temperature structure and composition prescribed by the fitted parameters. We include H$_2$, He, HCN, H$_2$O, CO, CO$_2$, CH$_4$, NH$_3$, and C$_2$H$_6$ in our nominal set of molecules. We perform additional tests that include DMS and DMDS. For absorbing species, we use the cross sections from HELIOS-K for H$_2$O \citep{ExoMol_H2O}, CO \citep{Hargreaves2019}, CO$_2$ \citep{ExoMol_CO2}, CH$_4$ \citep{HargreavesEtal2020apjsHitempCH4}, HCN \citep{Harris_HCN_2006}, NH$_3$ \citep{ExoMol_NH3}. The C$_2$H$_6$ cross-sections are computed from the HITRAN line lists \citep{Gordon2022_HITRAN}. For DMS and DMDS, where no line lists are available, we use the 278.15\,K cross-sections from the HITRAN database \citep{sharpe_DMS_2004}. The abundances of absorbing species are fitted as parameters, while H$_2$/He is treated as a filler gas making up the remainder for an atmosphere (with a Jupiter-like ratio He/H$_2$ = 0.157). 

Contrary to emission spectra, transmission spectra struggle to constrain gradients in the vertical temperature structure of sub-Neptunes. This motivates us to adopt the standard prescription of instead fitting only one temperature $T_p$ (the photospheric temperature) to be near-constant over the limited pressure range probed in transmission. To account for aerosol obfuscation, we use a gray cloud deck (fitting the pressure below which the atmosphere becomes opaque) and hazes following \citet{lecavelier_des_etangs_rayleigh_2008}. We do not include stellar contamination in our model, since the deeper analysis from \citetalias{Schmidt2025arXiv250118477S} found that its inclusion does not impact the inferred properties from the NIRISS+NIRSpec data. The longer wavelengths probed by MIRI are not sensitive to stellar contamination. We allow for three offsets between instruments/detectors relative to NIRISS/SOSS: one for the NRS1 detector of NIRSpec/G395H, one for the NRS2 detector of NIRSpec/G395H, and another for MIRI/LRS. We also fit the planet mass with a Gaussian prior on the value reported by \citet{Benneke2019ApJ...887L..14B}.

For each combination of model parameters, we optimize the value of the 10\,mbar radius until the best data-model match is achieved, iterating over both the hydrostatic equilibrium and radiative transfer steps.  The parameter space is sampled using multi-ellipsoid nested sampling as implemented in the \texttt{nestle} module \citep{skilling_nested_2004, skilling_nested_2006}. We compute models at a resolving power of 20,000, appropriate for the 100--200 resolving power of the data at hand. Each model is then convolved to the resolution of the observed spectrum for the likelihood evaluation.

\section{Molecular Abundances}

\renewcommand{\arraystretch}{1.2}
\begin{sidewaystable*}
    \centering
    \small
    \caption{Retrieved molecular abundances of prominent molecules in the atmosphere of K2-18\,b. In the SCARLET retrievals, the CO abundance posteriors fill the entire range of the prior, leading to no abundance constraints, although the molecule is included in the model (N/A in the table). Upper limits mark the 95th percentile of the posterior. The log-evidence of each model is reported.}
    \label{table:abundances}
    \begin{tabular}{ll|cccccc|ccc|c}
        \hline\hline
        Models & Dataset & $\log X_{\rm CH_4}$ & $\log X_{\rm CO_2}$ & $\log X_{\rm H_2O}$ & $\log X_{\rm NH_3}$ & $\log X_{\rm HCN}$ & $\log X_{\rm CO}$ & $\log X_{\rm DMS}$ & $\log X_{\rm DMDS}$ & $\log X_{\rm C_2H_6}$ & $\ln Z$ \\
        \hline
        \multicolumn{11}{c}{PLATON analyses} \\ 
        \hline
        Baseline                    & \texttt{JExoRES} & $-0.85_{-0.32}^{+0.20}$ & $-1.89_{-0.86}^{+0.49}$ &   $< -3.59$ & $< -5.28$ & $< -3.39$ & $< -3.3$ & \dots & \dots & \dots & 29035.03 \\
        With DMS/DMDS               & \texttt{JExoRES} & $-0.89_{-0.33}^{+0.22}$ & $-1.95_{-0.96}^{+0.51}$ &   $< -3.4$ & $< -5.05$ & $< -3.18$ & $< -3.63$ & $< -4.17$ & $< -4.34$ & \dots & 29034.72 \\
        With DMS/DMDS - No MIRI     & \texttt{JExoRES} & $-0.84_{-0.30}^{+0.19}$ & $-1.91_{-0.97}^{+0.50}$ &   $< -3.74$ & $< -5.12$ & $< -3.35$ & $< -3.4$ & $< -4.48$ & $< -4.93$ & \dots & 28820.43 \\
        With C$_2$H$_6$             & \texttt{JExoRES} & $-0.94_{-0.37}^{+0.26}$ & $-1.91_{-0.95}^{+0.52}$ &  $< -3.72$ & $< -5.27$ & $< -3.46$ & $< -3.65$ & \dots & \dots  & $< -1.05$ & 29035.69 \\
        With DMS/DMDS/C$_2$H$_6$    & \texttt{JExoRES} & $-0.92_{-0.34}^{+0.24}$ & $-1.97_{-0.97}^{+0.55}$ &   $< -3.56$ & $< -5.04$ & $< -3.29$ & $< -3.61$ & $< -4.37$ & $< -4.52$ & $< -1.14$ & 29035.08 \\
        \hline
        \multicolumn{11}{c}{SCARLET analyses} \\ 
        \hline
        Baseline                    & \texttt{exoTEDRF}     & $-1.25_{-0.73}^{+0.44}$ & $-3.60_{-3.69}^{+1.70}$ & $< -2.96$ & $< -3.66$ & $< -2.62$ & N/A & \dots & \dots & \dots & $-1935.975$ \\
        With DMS/DMDS               & \texttt{exoTEDRF}     & $-1.16_{-0.67}^{+0.35}$ & $ -4.83_{-4.22}^{+2.62}$ & $< -2.26$ & $< -3.32$ & $< -2.40$ & N/A & $< -3.36$ & $< -2.70$ & \dots & $-1933.338$ \\
        With C$_2$H$_6$             & \texttt{exoTEDRF}     & $-1.2 _{ -0.71 }^{+ 0.42 }$ & $-3.55 _{ -4.39 }^{+ 1.76 }$ & $< -2.69$ & $< -3.53$ & $< -2.6$ & N/A & \dots & \dots & $N/A$ & $-1935.819$\\
        With DMS/DMDS/C$_2$H$_6$    & \texttt{exoTEDRF}     & $-1.11 _{ -0.59 }^{+ 0.33 }$ & $-4.93 _{ -4.24 }^{+ 2.74 }$ & $< -2.33$ & $< -3.49$ & $< -2.5$ & N/A & $< -3.26$& $<-2.74$ & N/A & $-1933.124$ \\
        \\
        Baseline                    & \texttt{exoTEDRF+SPARTA}  & $-1.61_{-0.72}^{+0.64}$ & $-1.66_{-1.50}^{+0.46}$ & $< -3.94$ & $< -4.60$ & $< -4.68$ & N/A & \dots & \dots & \dots & $-1883.560$ \\
        With DMS/DMDS               & \texttt{exoTEDRF+SPARTA}  & $-1.04_{-0.53}^{+ 0.32}$ & $-2.13_{-3.98}^{+0.77}$ & $ < -2.55$ & $< -4.08$ & $< -3.07$ & N/A & $< -2.87$ & $< -4.07$ & \dots & $-1881.242$ \\
        With C$_2$H$_6$             & \texttt{exoTEDRF+SPARTA}         & $-1.57_{-0.67}^{+ 0.60}$ & $-1.61 _{ -1.54 }^{+ 0.45 }$ & $< -3.73$ & $< -4.56$ & $< -4.41$ & N/A & \dots & \dots & N/A & $-1883.651$ \\
        With DMS/DMDS/C$_2$H$_6$    & \texttt{exoTEDRF+SPARTA}         & $-1.04 _{ -0.57 }^{+ 0.31 }$ & $-1.99 _{ -4.53 }^{+ 0.66 }$ & $< -2.72$ & $< -4.11$ & $< -2.95$ & N/A & $< -2.79$ & $< -4.12$ & $N/A$ & $-1880.930 $ \\ 
        \hline
         \hline
    \end{tabular}
\end{sidewaystable*}

\end{appendix}

\end{document}